\documentclass[]{pasj01}
\draft

\newcommand\sun{\odot}%
\newcommand\degr{\arcdeg}%

\begin{document} 
\Received{}
\Accepted{}

\title{Radial Velocity Measurements of an Orbiting Star Around Sgr\,A*}

\author{Shogo \textsc{Nishiyama}\altaffilmark{1,*}%
}
\altaffiltext{1}{Miyagi University of Education, Sendai 980-0845, Japan}
\email{shogo-n@staff.miyakyo-u.ac.jp}

\author{Hiromi \textsc{Saida}\altaffilmark{2}}
\altaffiltext{2}{Daido University, Nagoya 457-8530, Japan}

\author{Yohsuke \textsc{Takamori}\altaffilmark{3}}
\altaffiltext{3}{National Institute of Technology, Wakayama College, Wakayama 644-0023, Japan}

\author{Masaaki \textsc{Takahashi}\altaffilmark{4}}
\altaffiltext{4}{Aichi University of Education, Aichi 448-8542, Japan}

\author{Rainer \textsc{Sch\"{o}del}\altaffilmark{5}}
\altaffiltext{5}{Instituto de Astrof\'{i}sica de Andaluc\'{i}a (IAA)-CSIC, 18008 Granada, Spain}

\author{Francisco \textsc{Najarro}\altaffilmark{6}}
\altaffiltext{6}{Centro de Astrobiolog\'{i}a (CSIC/INTA), 28850 Madrid, Spain}

\author{Satoshi \textsc{Hamano}\altaffilmark{7}}
\altaffiltext{7}{Kyoto Sangyo University, Kyoto 603-8555, Japan}

\author{Masashi \textsc{Omiya}\altaffilmark{8}}
\altaffiltext{8}{National Astronomical Observatory of Japan, Tokyo, 181-8588, Japan}

\author{Motohide \textsc{Tamura}\altaffilmark{8,9}}
\altaffiltext{9}{The University of Tokyo, Tokyo, 113-0033, Japan}

\author{Mizuki \textsc{Takahashi}\altaffilmark{10}}
\altaffiltext{10}{Tohoku University, Sendai 980-8578, Japan}

\author{Haruka \textsc{Gorin}\altaffilmark{1}}

\author{Schun \textsc{Nagatomo}\altaffilmark{11}}
\author{Tetsuya \textsc{Nagata}\altaffilmark{11}}
\altaffiltext{11}{Kyoto University, Kyoto 606-8502, Japan}

\KeyWords{black hole physics --- relativistic processes --- instrumentation: high angular resolution --- instrumentation: spectrographs --- methods: observational --- techniques: radial velocities --- Galaxy: center --- infrared: stars} 

\maketitle

\begin{abstract}
During the next closest approach of the orbiting star S2/S0-2 
to the Galactic supermassive black hole (SMBH),
it is estimated that RV uncertainties of $\sim 10\,$km/s allow us to detect 
post-Newtonian effects throughout 2018.
To evaluate an achievable uncertainty in RV and its stability,
we have carried out near-infrared, high resolution 
($R \sim 20,000$) spectroscopic monitoring observations of S2
using the Subaru telescope and the near-infrared spectrograph IRCS from 2014 to 2016.
The Br-$\gamma$ absorption lines are used to determine the RVs of S2.
The RVs we obtained are 497\,km/s, 877\,km/s, and 1108\,km/s
in 2014, 2015, and 2016, respectively.
The statistical uncertainties are derived using the jackknife analysis.
The wavelength calibrations in our three-year monitoring are stable:
short-term (hours to days) uncertainties in RVs are $\lesssim 0.5$\,km/s,
and a long-term (three years) uncertainty is 1.2\,km/s.
The uncertainties from different smoothing parameter,
and from the partial exclusion of the spectra, are found to be a few km/s.
The final results using the Br-$\gamma$ line are
$497 \pm 17 \mathrm{(stat.)} \pm 3 \mathrm{(sys.)}$\,km/s in 2014,
$877 \pm 15 \mathrm{(stat.)} \pm 4 \mathrm{(sys.)}$\,km/s in 2015, and
$1108 \pm 12 \mathrm{(stat.)} \pm 4 \mathrm{(sys.)}$\,km/s in 2016.
When we use two He\,I lines at 2.113$\,\mu$m in addition to Br-$\gamma$,
the mean RVs are 513\,km/s and 1114\,km/s for 2014 and 2016, respectively.
The standard errors of the mean are 16.2\,km/s (2014) and 5.4\,km/s (2016),
confirming the reliability of our measurements.
The difference between the RVs estimated by Newtonian mechanics and general relativity 
will reach about 200 km/s near the next pericenter passage in 2018.
Therefore our RV uncertainties of $\approx 13 - 17\,$km/s with Subaru enable us to detect 
the general relativistic effects in the RV measurements with more than $10 \,\sigma$ in 2018.

\end{abstract}

\section{Introduction}
\label{sec:intro}

At the center of our Galaxy,
a dark mass of $\sim 4 \times 10^6 M_{\sun}$ is likely to be associated
with the compact radio source Sgr\,A*.
In the immediate vicinity of Sgr\,A*,
a number of rapidly orbiting stars (called S-stars) has been detected
(e.g., \cite{Eckart96Nat, 96GenzelApJ, Ghez98ApJ}),
and their orbits have been determined 
(e.g., \cite{Schodel02Nat, Ghez05ApJ}).
The motions of the stars around Sgr\,A* have given a lower limit
of the mass density inside their pericenter,
which provides one of the most compelling cases so far 
for the existence of supermassive black holes
(SMBHs; \cite{Boehle16ApJ,17GillessenApJ, Parsa17Aph}, for most recent works).

The stellar system around Sgr\,A* provides an unique test bed
for probing the strong gravitational field around a SMBH.
The orbiting S-stars can be regarded as test particles
moving in the gravitational field generated by the Galactic SMBH,
and particularly important is the star 
S2 (in the VLT nomenclature) or S0-2 (in the Keck nomenclature).
S2/S0-2 is orbiting Sgr\,A* in $\approx 16$\,yr,
which is one of the shortest periods among the orbiting stars,
and has a large orbital eccentricity of $e \approx 0.88 - 0.89$
\citep{Boehle16ApJ,17GillessenApJ, Parsa17Aph}.
With a magnitude of $K_{S} \sim 14$, 
S2 is the brightest of the short-period stars.
These mean that S2 is an ideal target to 
observe the strong gravitational field around the SMBH,
and we expect to detect the post-Newtonian (PN) effects
(i.e., deviation from the Newtonian gravity)
with current telescopes.
The closest approach of S2 to Sgr\,A*  is expected to be at
$2018.29 - 2018.59$ \citep{Boehle16ApJ, 17GillessenApJ, Parsa17Aph} .
General relativistic (GR) effects are strongest near the pericenter of the orbit,
where the pericenter distance is only about 120\,AU, 
and the speed of S2 reaches a few \% of the speed of light.
The observations of the orbital motion and light trajectories of S2 in 2018
therefore provide an opportunity to test unobserved predictions of GR around the SMBH,
and to understand the nature of gravity.

The directly observable quantities of S2 dynamics are
its positions in the sky ($\alpha$ and $\delta$) in astrometric measurements,
and radial velocities (RVs) in spectroscopic measurements.
Given the current measurement precision,
establishing an accurate astrometric reference frame is the greatest challenge,
and identifying sources of systematic uncertainties is the ongoing study
(see, e.g., \cite{Gillessen09PM, Yelda10GC, Plewa15MNRAS}).
On the other hand, 
when we obtain new spectroscopic data, 
RV simply refers to the Local Standard of Rest (LSR).
More direct comparison of observed RVs with theoretical prediction is possible
with respect to the astrometric observations.

We have been focusing on the spectroscopic measurements of S2.
Note that, in the context of both of general and special relativities, 
the redshift of observed photons, $z$, 
and the RV of S2,  $v_{\mathrm{S2}}$, are in a complicated non-linear relation.
Our observable quantity is not exactly $v_{\mathrm{S2}}$, but
\begin{equation}
	z = \frac{\lambda_{\mathrm{obs}}}{\lambda_{\mathrm{S2}}} -1 \neq \frac{v_{\mathrm{S2}}}{c}
	\label{eq:eq1}
\end{equation}
where $\lambda_{\mathrm{S2}}$  and $\lambda_{\mathrm{obs}}$ are wavelengths 
emitted from S2 and measured in an observation, respectively,
 and $c$ is the speed of light. 
However, following the traditional nomenclature, 
we show our observed value $cz$ in the unit of km/s 
and call this value not the redshift but the ``radial velocity (RV)".

The RV measurements of S2 near its pericenter passage
allow us to detect the PN effects
\citep{Zucker06ApJ, Iorio11MNRAS, Angelil10ApJ, Angelil11ASPC, Zhang15ApJ, Yu16ApJ, Zhang17ApJ, Grould17IAUS, Hees17arXiv}.
In the complicated relation between $cz$ and $v_{\mathrm{S2}}$, 
the kinematic Doppler-shift and the gravitational redshift appear as
the strongest effects and they show comparable amplitudes.
Those two effects are estimated to be as large as 
about 200 km/s at the pericenter passage of S2, 
and current instruments are capable of detecting 
at least these RV shifts in the spectroscopic measurements \citep{Zucker06ApJ}.

However, we have significant constraints for observations of the S-stars \citep{Schodel15IAU}.
The interstellar extinction toward the Galactic center is extreme,
more than $A_V = 30$\,mag.
It means that we cannot observe stars near Sgr\,A* in the optical wavelength.
In addition, stellar number density is so high that
without an adaptive optics (AO) system,
our observations are limited to relatively bright magnitudes
due to the crowding.
Therefore NIR instruments with AO are crucial for observational studies to resolve the S-stars.

In the past, most of the spectroscopic observations of S2 were carried out
with medium spectral resolution instruments.
The instruments used are NACO/VLT 
(long slit spectroscopy with $R \sim 1,400$; \cite{03EisenhauerApJ}),
SPPIFI/VLT (IFU with $R \sim 3,500$; \cite{03EisenhauerApJ}),
SINFONI/VLT (IFU with $R \sim 1,500 - 4,000$; \cite{Gillessen09PM,17GillessenApJ}),
NIRSPEC/Keck (long slit spectroscopy with $R \sim 2,600$; \cite{Ghez08PM}),
NIRC2/Keck (long slit spectroscopy with $R \sim 4,000$; \cite{Ghez08PM}),
OSIRIS/Keck (IFU $R \sim 3,600$; \cite{Ghez08PM, Meyer12Sci, Boehle16ApJ}).
The mean of the RV uncertainties since 2010 is $\sim 34$\,km/s,
with the best of 15\,km/s.
Hence, the two leading PN effects 
(the transverse Doppler and gravitational redshift) can be detected
with uncertainties on the order of $5 \sigma$ through the observations in 2018,
if the astrometrically measured positions were known with infinite precision.
However, it is still difficult to monitor the time evolution
of the PN effects with an uncertainty of more than $\sim 30\,$km/s.

We focus on the spectroscopic observations of S2 to determine
the time variation of its RVs, as precise as possible.
We have carried out near-infrared (NIR), high spectral resolution spectroscopy
using the Subaru telescope.
The aim of this paper is to evaluate the possibilities of 
more accurate and stable RV measurements of S2 than the past ones.
Below, we show results of our spectroscopic monitoring observations of S2
from 2014 to 2016.
We also discuss shortly the expectation for the detection of 
the PN effects in the spectroscopic measurements during 2018.

\section{Data and Observation}

We have conducted out NIR spectroscopic observations
using the Subaru telescope \citep{Iye04Subaru}
and IRCS \citep{00KobayashiSPIE}.
The observations were carried out during the nights of
18 May 2014, 20 Aug 2015, and $17 - 18$ May 2016 (Table \ref{Tab:Obs}).
The IRCS echelle mode provides a spectral resolution of 
$\lambda / \Delta \lambda \approx 20,000$ in the $K$ band.
The slit length and width are 5\farcs17 and 0\farcs14, respectively.
We took spectra in the $K+$ setting, to include Br-$\gamma$ absorption line at 2.16612\,$\mu$m,
and He\,I  absorption lines at 2.11260\,$\mu$m and 2.11378\,$\mu$m\footnote
{Vacuum wavelengths from http://www.pa.uky.edu/\textasciitilde peter/newpage/}
(Table \ref{Tab:IRCS_EchelleKp}).

\begin{table}[htb]
\begin{center}
\caption{Summary of observations. \label{Tab:Obs}} 
\begin{tabular}{ccccccc}
\hline
\hline
Date & Setting & IT$^{\mathrm{a}}$ & $N_{\mathrm{frame}}$$^{\mathrm{b}}$ & $N_{\mathrm{used}}$$^{\mathrm{c}}$ & slit angle$^{\mathrm{d}}$ & AO$^{\mathrm{e}}$  \\
(UTC) & & [sec] & & & [degree]\\
\hline
2014 May 19 & $K+$ & 300 & 32 & 30 & 8 & LGS\\
2015 Aug 21 & $K+$ & 300 & 24 & 24 & 8 & NGS\\
2016 May $17-18$ & $K+$ & 300 & 48 & 44 & 8, 128 & LGS \\
\hline
\end{tabular}\\
\end{center}
(a) Integration time for each exposure.\\
(b)The number of frame taken in the night(s).\\
(c)The number of frame used in data analysis.\\
(d)The angular offset measured from North to East, counterclockwise direction.\\
(e)The guide star of the AO system. The ``LGS" mode uses the laser guide star system, and the ``NGS" mode uses only a natural guide star.\\
\end{table}

\begin{table}[htb]
\begin{center}
\caption{Wavelength coverage of the IRCS echelle $K+$ mode and 
	the number of OH emission lines used for wavelength calibration. 
	\label{Tab:IRCS_EchelleKp}} 
\begin{tabular}{cccc}
\hline
\hline
Order & Wavelength coverage [$\mu$m] & Dispersion [/pixel] & $N_{\mathrm{OH}}~^{\mathrm{a}}$ \\
\hline
23 & $2.4456 - 2.5038$ & 0.569 & - \\
24 & $2.3437 - 2.3997$ & 0.547 & - \\
25 & $2.2500 - 2.3039$ & 0.527 & 6 \\
26 & $2.1634 - 2.2154$ & 0.508 & 7 \\
27 & $2.0833 - 2.1335$ & 0.491 & 14 \\
28 & $2.0089 - 2.0575$ & 0.475 & 5 \\
29 & $1.9397 - 1.9867$ & 0.460 & 9 \\
\hline
\end{tabular}
\end{center}
(a) The maximum number of the atmospheric OH line used for wavelength calibration.
\end{table}

On the nights in 2014, 2015, and the first night in 2016,
the position angle of the slit was set to be $\approx 8\degr$
\footnote{
The position angle of the slit is defined as the angular offset in degrees 
relative to the north celestial pole. 
The angle is measured from North to East, counterclockwise direction.}.
In this setting, a bright star IRS 16NW
(Ofpe/WN9, $K=10.1$; \cite{Paumard06}) is observable as well as S2.
On the second night in 2016, the position angle was set to be 128\degr,
and a bright star IRS 29N (WC9, $K=10.0$; \cite{Paumard06}) is on the slit simultaneously.
The bright stars are used to trace the positions of the S2's spectra
on the array in the data reduction procedure.

The adaptive optics (AO) system on Subaru, AO188 \citep{Hayano08SPIE, Hayano10SPIE}, 
was used in our observations.
The laser guide star (LGS) was propagated at the center of our field in the 2014 and 2016 runs.
Since the LGS system did not work well in 2015,
instead, an $R = 13.8$\,mag star USNO 0600-28577051 was used
as a natural guide star.
USNO 0600-28577051 was used as a tip-tilt guide star in 2014 and 2016.

Thirty-two exposures were obtained on 18 May 2014.
Due to thin clouds, we cannot find spectra of S2 for two exposures,
and we thus use 30 exposures in the following analysis.
In 2015 and 2016, 24 and 48 exposures were obtained, respectively.
In 2016, S2 spectra cannot be clearly seen in four exposures.
The AO guide star was lost in the exposures due to thin clouds coming.
The exposure time is always 300 sec from 2014 to 2016.

We have observed standard stars during our runs.
HD 152521 (A0-1V), HD 171296 (A0V), or HD 183997 (A0IV-V), 
was observed once or twice per night.
Since the region around S2 is very crowded,
we have observed a dark cloud located at a few arcmin northwest 
from S2, to obtain sky measurements.
During the sky observations, we confirmed that
no object is included in the slit position.

\section{Data Reduction and Analysis}

The reduction procedure includes dark subtraction, flat-fielding,
bad pixel correction, cosmic-ray removal, sky subtraction, 
spectrum extraction, wavelength calibration, 
telluric correction, and spectrum continuum fitting.
Flat field images were obtained through observations of
a continuum source.
The sky field, a few arcmin northwest from S2,
was observed once or twice per night,
and time differences between the sky and objects are as large as 2\,hrs.
This is longer than the typical variability of the sky in the NIR wavelength.
We thus scaled the sky images to subtract atmospheric OH emission lines
as cleanly as possible.


We have extracted S2 spectra using the IRAF \textit{apall} task in the \textit{echelle} package.
The S2 spectra are faint, and thus bright spectra of IRS 16NW or IRS 29N
are used to trace the positions of S2 spectra,
in the assumption that the spectra of S2 and that of the bright stars run parallel to each other.
Background subtraction is carried out in the \textit{apall} task.
For each exposure, we selected some blank regions around S2 by eye using ``aperture editor".
The median value of the selected regions are used for the background subtraction.

\subsection{Wavelength calibration}

The wavelength calibration is carried out for the extracted spectra.
The wavelength solutions are obtained by identifying the atmospheric OH emission lines,
or atmospheric absorption lines.
At the end of our observations, we took arc lamp exposures for the wavelength calibration.
However, the calibration with the arc lamp resulted in large uncertainties, probably because
they were taken at different time, and the number of lines is small,
only 12 lines in the echelle orders from 25 to 29. 

We also tried to obtain the wavelength solutions by using the OH lines in the sky frames.
However, we have observed the sky fields only once or twice per night to maximize 
the exposure time of S2,
leading to a time difference of $\lesssim 2$ hrs between the sky and S2 observations.
This is likely to be a main reason of the low accuracy of the wavelength calibration
using the OH lines in the sky frames.
When we use the OH lines as the wavelength calibration,
telluric-corrected spectra show many residuals 
at the position of the atmospheric absorption lines.

We therefore decided to determine the wavelength calibration
directly from the science data in the following way:
We have taken S2 spectra in a traditional A position $-$ B position manner.
For the wavelength calibration of S2 spectra in an A position exposure, 
we use OH lines in spectra at the same coordinates on the array 
in the B-position exposure.
Similarly, for the calibration of B-position spectra,
OH lines in the A-position exposure is used.
As a result, it becomes possible to use the OH emission lines at the same coordinates on the array, 
taken at almost the same time as the spectra of S2, for the wavelength calibration.

We have used as many OH lines as possible for the wavelength calibration.
The number of the OH lines detected in our spectra are 41 in total,
in the echelle orders from 25 to 29 (Table \ref{Tab:IRCS_EchelleKp}).
In some exposures, the S/N ratio of some OH lines was too low to be useful for the calibration  
because the OH lines were contaminated by weak and noisy stellar continuum spectra,
and thus the smaller number of the OH lines was used.

\subsection{Telluric Correction using Standard Stars and Continuum Fitting}

We observed standard star(s) once or twice per night.
Since isolated standard stars are selected, 
we carried out a traditional A$-$B and B$-$A reduction.
The standard stars are bright enough to obtain continuum spectra with a high S/N ratio,
and the atmospheric absorption lines are used for the wavelength calibration.
The number of the absorption lines used for the calibration is 269
in the echelle orders from 25 to 29.
Prior to division of S2 spectra by the standard star spectra,
the Br-$\gamma$ absorption line was removed from the standard star spectra
by fitting the absorption profile with a Moffat function using the IRAF \textit{splot} task.
The S2 spectra were then divided by the standard star spectra
using the \textit{telluric} task.

\begin{figure}
 \begin{center}
   \includegraphics[width=0.8\textwidth,angle=0]{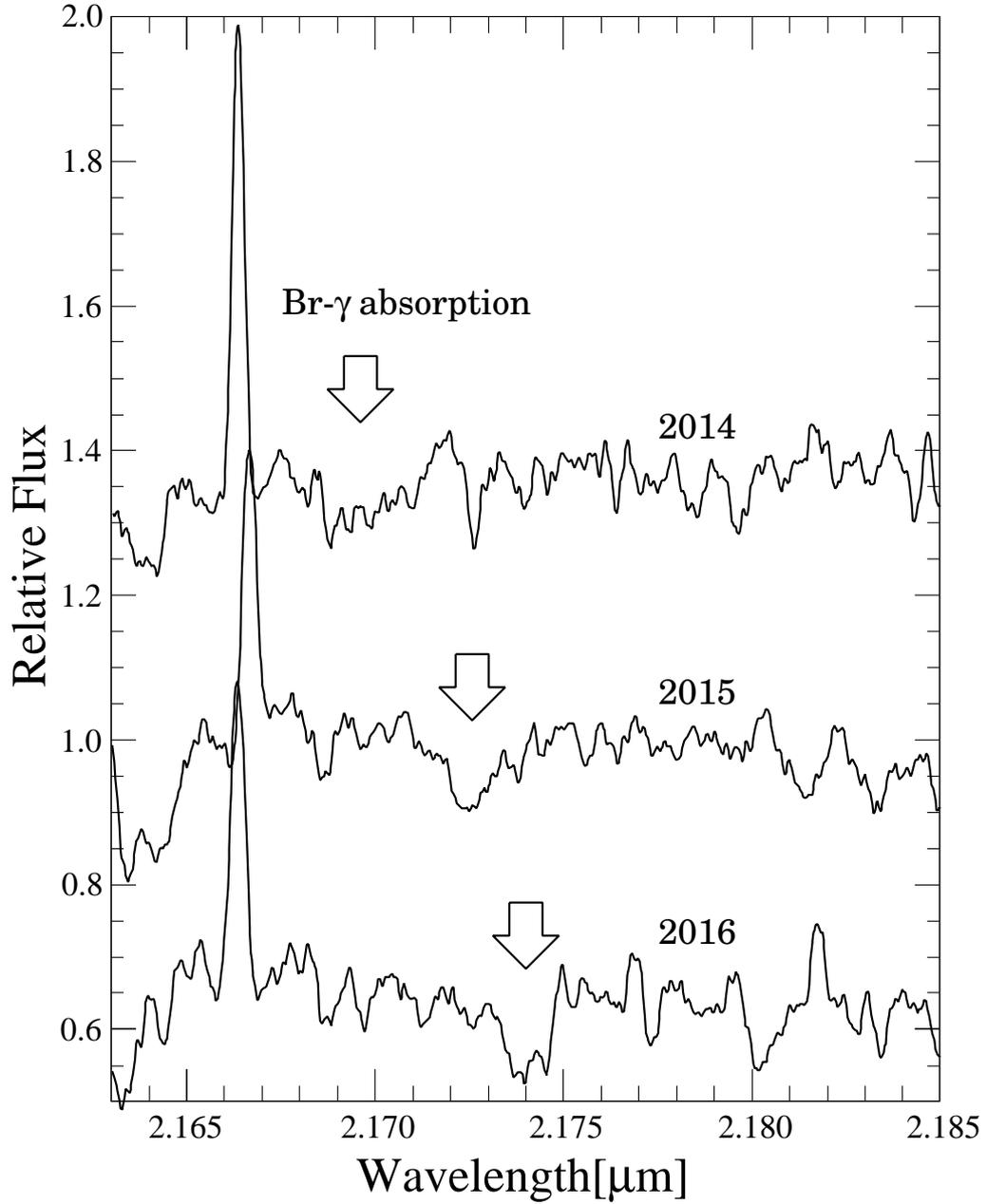}
 \end{center}
\caption{
    Spectra including Br-$\gamma$ absorption and emission lines of S2 
    in May 2014 (top), Aug 2015 (middle), and May 2016 (bottom).
    The positions of the Br-$\gamma$ absorption features are shown by white arrows.
    In all cases the smoothing parameters was chosen as $s = 9$.
    The Br-$\gamma$ emission lines from ambient gas are seen 
    at $2.166 \mu$\,m $- 2.167 \mu$\,m.
    The emission line was not removed to improve the line detection in low S/N spectra.
    The spectra are not corrected to have a LSR wavelength.
}\label{fig:specBr-g141516}
\end{figure}

We have carried out continuum fit with the IRAF \textit{continuum} task
for the telluric-corrected spectrum of each exposure.
After the fitting, the spectra are median combined with the \textit{scombine} task.
The combined spectra around Br-$\gamma$ emission are shown in Fig. \ref{fig:specBr-g141516}.

\subsection{Correction to LSR}

To obtain RVs in the LSR reference frame,
we need to take into account the following motions:
the rotation of the Earth; the orbital motion around the Sun;
and the Sun's peculiar motion with respect to the LSR.
The amount of the RV correction,  $\Delta \mathrm{RV}_{\mathrm{LSR}}$, 
was calculated using the IRAF \textit{rvcorrect} task.
The calculated LSR velocities are shown in Table \ref{Tab:LSR}.
We have used the mean of the values, the start and the end of the observations,
for the correction of the motions:
$+24.6$ km/s for 2014;
$-15.7$ km/s for 2015;
and $+24.5$ km/s for 2016.

\begin{table}[htb]
\begin{center}
\caption{RV correction with respect to LSR. \label{Tab:LSR}} 
\begin{tabular}{clclclclclclll}
\hline
\hline
Date & start IT$^{\mathrm{a}}$ & End IT$^{\mathrm{b}}$
& $\Delta$RV$_{\mathrm{LSR},1}$$^{\mathrm{c}}$ 
& $\Delta$RV$_{\mathrm{LSR},2}$$^{\mathrm{d}}$ \\
 (UTC)  & (UTC) & (UTC) & [km/s] \\
\hline
2014 May 19 & 10:59:47.82 &14:48:45.74 & $+24.74$ & $+24.41$ \\
2015 Aug 21 & 06:36:39.99 & 08:46:39.27 & $ -15.56$ & $-15.78$ \\ 
2016 May 17  & 10:47:22.88 & 14:22:37.34 & $+24.97$ & $+24.56$ \\
2016 May 18  & 13:06:56.76 & 14:27:09.29 & $+24.26$ & $+24.11$ \\
\hline
\end{tabular}
\end{center}
(a) The time of the start of integration for the 1st frame of S2 observations.\\
(b) The time of the end of integration for the last frame of S2 observations.\\
(c) The local standard of rest velocity at the time of start IT.\\
(d) The local standard of rest velocity at the time of end IT.\\
\end{table}

\section{Measured Radial Velocities and their Uncertainties}
\label{sec:RVandUncertainty}

\begin{figure}[!ht]
 \begin{center}
      \includegraphics[width=0.6\textwidth,angle=0]{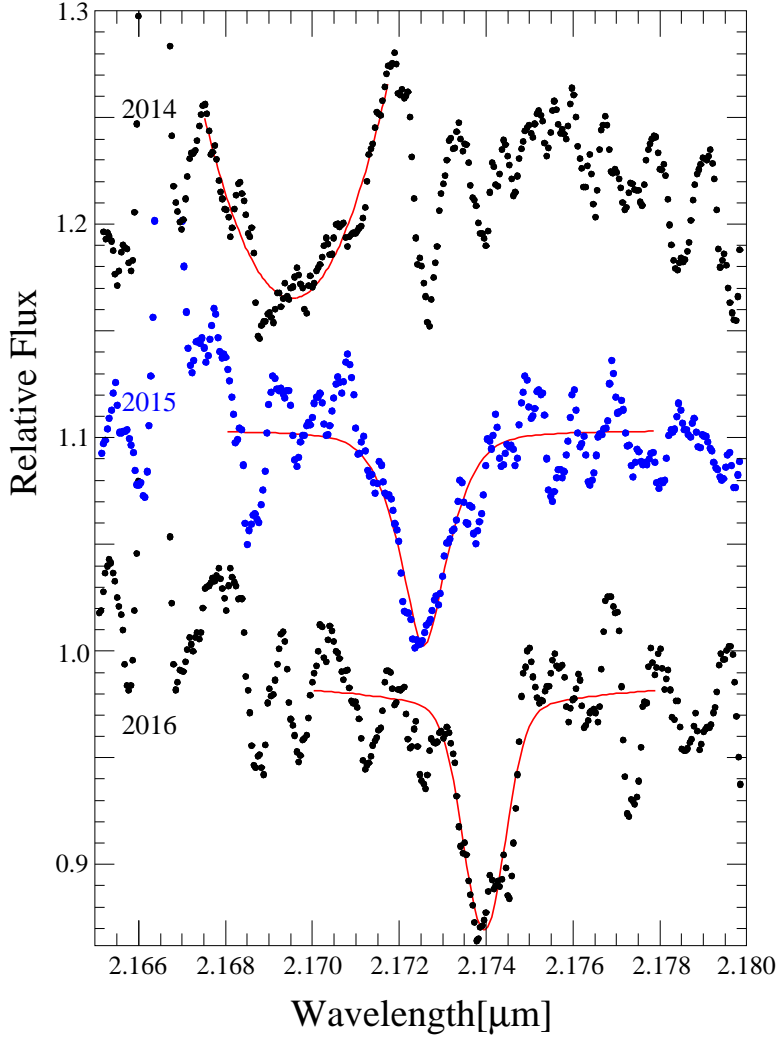}
     \end{center}
    \caption{
	Combined spectra around the Br-$\gamma$ absorption line
	for 2014 (top), 2015 (middle, blue dots), and 2016 (bottom).
	The Br-$\gamma$ profiles are fit with a Moffat function 
	for 2015 and 2016 (red lines).
	The 2014 spectrum is fit with a Gaussian function.
	The smoothing parameter for the shown spectra is $s = 11$.
    }
    \label{fig:specBr-g141516Allfit}
\end{figure}

RVs of S2 are determined for each spectrum on the basis of 
the location of the Br-$\gamma$ absorption line.
The combined spectra around the Br-$\gamma$ absorption line are given 
in Fig. \ref{fig:specBr-g141516Allfit},
which shows how the Br-$\gamma$ line has shifted from 2014 to 2016.
To determine the central wavelength of the lines, 
a combination of Gaussian plus Lorentzian functions (i.e., Moffat function) is used
to fit the Br-$\gamma$ line profiles.
The peaks of both functions are set to have the same value.
Probably due to a noisy continuum and wide profile,
the 2014 spectrum cannot be fit with a Moffat function.
Hence it is fit with a Gaussian function.
The wavelength of the best fit peak is compared to the rest wavelength, $2.166120\,\mu$m,
to determine RV$_{\mathrm{LSR}}$,
with the correction of $\Delta \mathrm{RV}_{\mathrm{LSR}}$.

\subsection{Uncertainty from Spectrum Smoothing}

Due to the faintness of S2 and confusion of unresolved sources around it,
the S/N ratios of the full resolution S2 spectra are only about 15.
We have thus used ``smoothed" spectra to determine the peak wavelengths
of the lines.
The smoothing parameter, $s$, represents the box size in pixels
used for the spectrum smoothing.
The mean flux of the consecutive $s$ pixels is adopted as
the flux at the mean wavelength $\lambda$ of the $s$ pixels.

\begin{figure}[!ht]
 \begin{center}
   \includegraphics[width=0.6\textwidth,angle=0]{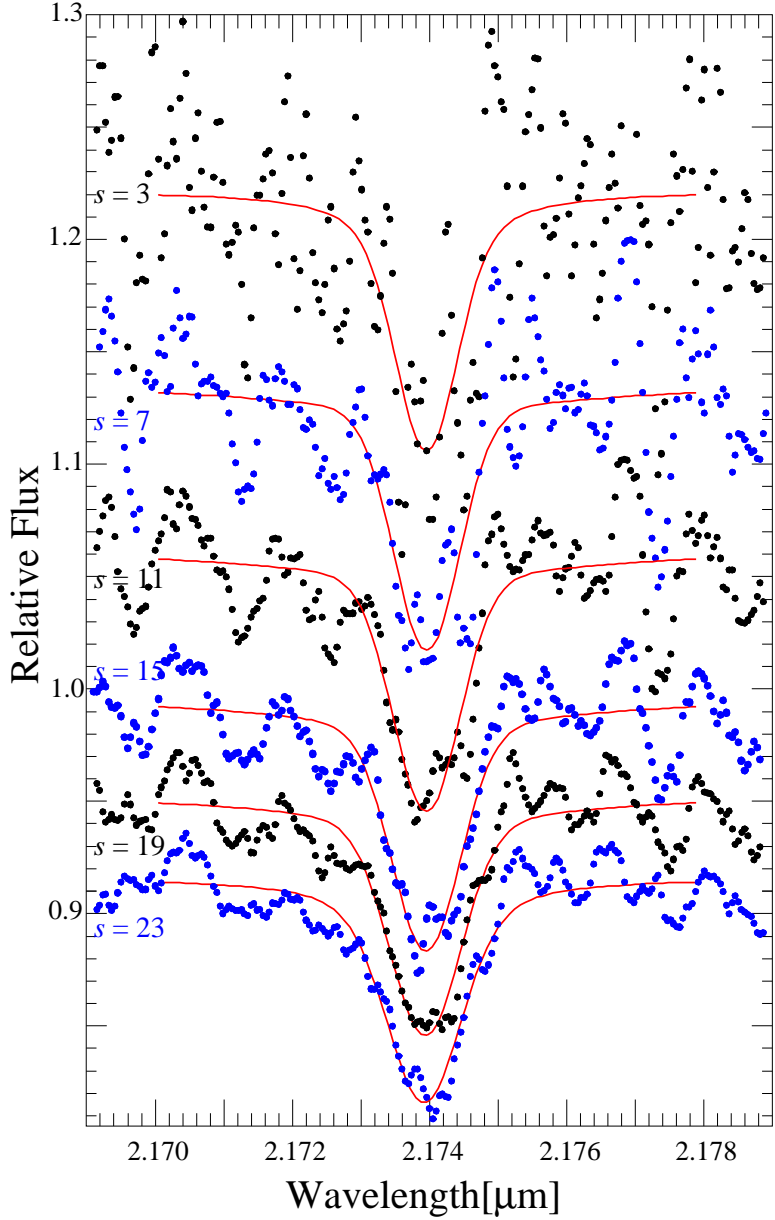}
 \end{center}
 \caption{
   Combined 2016 spectra around the Br-$\gamma$ absorption line
   with different smoothing parameters.
   From top to bottom, $s = 3$, 7, 11, 15, 19, and 23.
   The Br-$\gamma$ profiles are fit with a Moffat function (red lines).
 }
 \label{fig:specBr-g16Smooth}
\end{figure}

Fig. \ref{fig:specBr-g16Smooth} represents spectra in 2016 around Br-$\gamma$
for different smoothing parameters.
As one can see, spectra with small $s$ are noisy, 
but the peak wavelengths obtained by fitting the profiles
with different $s$ parameters are almost the same.

Table \ref{Tab:Brg_16_spar} gives the peak wavelengths of the Br-$\gamma$ absorption line
and resultant RV$_{\mathrm{LSR}}$ for different smoothing parameters.
In 2014, the standard deviation of the peak wavelengths 
for the different smoothing parameters $(s = 3 - 23)$ is 
$0.00011 \mu$\,m, which corresponds to a RV of $1.5$\,km/s.
Similarly, the standard deviations of the RV$_{\mathrm{LSR}}$ are 2.7\,km/s and 1.9\,km/s
for 2015 and 2016, respectively.
These are significantly smaller than the statistical uncertainties derived below.

\begin{table}[!ht]
\begin{center}
\caption{Wavelength of Br-$\gamma$ and RV$_{\mathrm{LSR}}$ for combined spectra,
with different smoothing parameters. \label{Tab:Brg_16_spar}} 
\begin{tabular}{ccccccc}
\hline
\hline
 & \multicolumn{2}{c}{2014}  & \multicolumn{2}{c}{2015} & \multicolumn{2}{c}{2016}  \\ \hline
$s$ & $\lambda$ [$\mu$\,m] & RV$_{\mathrm{LSR}}$ [km/s] 
& $\lambda$ [$\mu$\,m] & RV$_{\mathrm{LSR}}$ [km/s] & $\lambda$ [$\mu$\,m] & RV$_{\mathrm{LSR}}$ [km/s]\\ \hline
 3 & 2.169546  & 498.8 & 2.172554 &  874.8 & 2.173956  &  1109.0 \\
 5 & 2.169539  & 497.8 & 2.172557 &  875.2 & 2.173961  &  1109.7 \\
 7 & 2.169533  & 497.0 & 2.172559 &  875.5 & 2.173959  &  1109.5 \\
 9 & 2.169537  & 497.5 & 2.172563 &  876.0 & 2.173957  &  1109.2 \\
11 & 2.169536  & 497.3 & 2.172567 &  876.5 & 2.173955  &  1108.9 \\
13 & 2.169533  & 496.9 & 2.172573 &  877.4 & 2.173952  &  1108.5 \\
15 & 2.169529  & 496.4 & 2.172579 &  878.2 & 2.173948  &  1107.9 \\
17 & 2.169526  & 496.0 & 2.172586 &  879.2 & 2.173943  &  1107.2 \\
19 & 2.169521  & 495.2 & 2.172595 &  880.4 & 2.173937  &  1106.4 \\
21 & 2.169515  & 494.4 & 2.172603 &  881.5 & 2.173929  &  1105.2 \\
23 & 2.169509  & 493.6 & 2.172611 &  882.7 & 2.173919  &  1103.9 \\ \hline
mean & 2.169529 & 496.4 & 2.172577 & 877.9 & 2.173946 & 1107.8 \\
$\sigma$ & 0.000011 & 1.5 & 0.000020 & 2.7 & 0.000014 & 1.9 \\
\hline
\end{tabular}
\end{center}
\end{table}

\subsection{Jackknife Analysis of RV Uncertainty}

\begin{figure}[!h]
  \begin{center}
    \includegraphics[width=0.7\textwidth,angle=0]{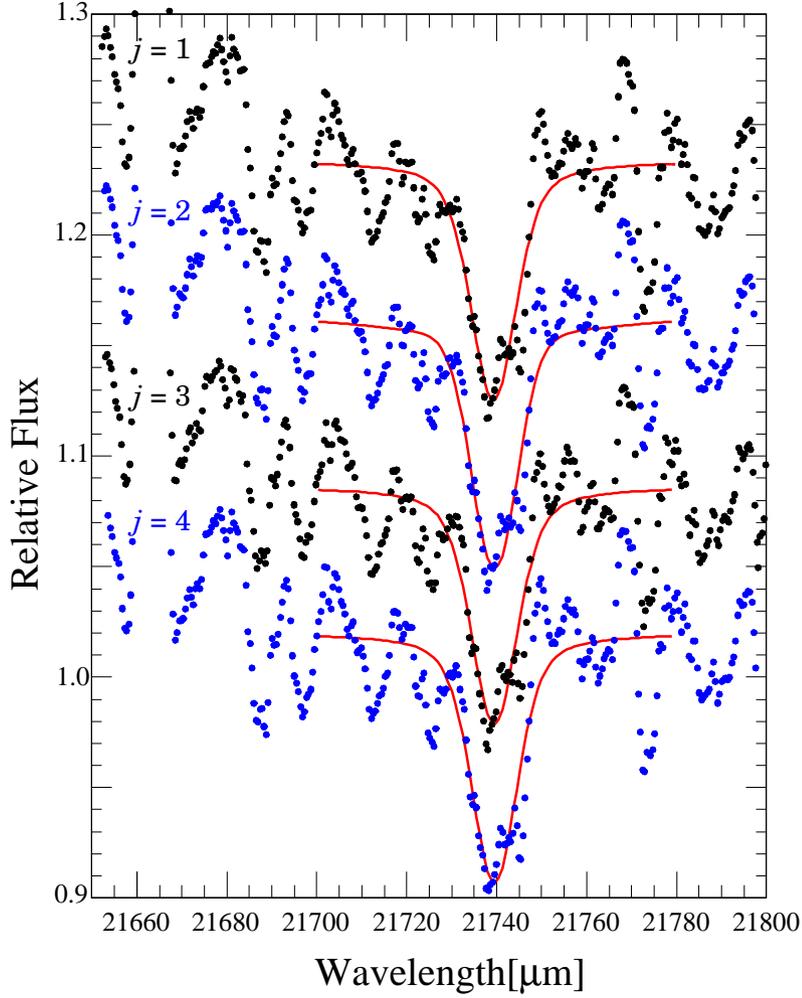}
  \end{center}
  \caption{
	Jackknife partial spectra around the Br-$\gamma$ absorption line for 2016.
	From top to bottom, they are spectra for the 1st, 2nd, 3rd, and 4th partial data sets.
	The Br-$\gamma$ profiles are fit with a Moffat function (red lines).
	The smoothing parameter for the shown spectra is  $s = 11$.
  }
  \label{fig:specBr-g16JK}
\end{figure}

We have used a jackknife analysis to estimate RV statistical uncertainties.
In the case of the 2016 data sets, 44 exposures were obtained.
The $j$th partial data set consists of 43 exposures without the $j$th exposure.
These 43 spectra were median combined, 
leading to the $j$th spectrum of S2.
In total, we obtain 44 partial-data spectra of S2.
The central wavelengths of the Br-$\gamma$ lines were obtained
by fitting the profiles with a combined function of Gaussian and Lorentzian.
The jackknife uncertainty, $\sigma_{\mathrm{JK}}$, is given by
\begin{equation}
	\sigma_{\mathrm{JK}} = \sqrt{ \frac{1}{N(N-1)} \sum_{j=1}^{N} (\alpha^*_j - \alpha^*)^2},
\end{equation}
where 
 \begin{equation}
	 \alpha^*_j = N \alpha - (N-1) \alpha_j,
\end{equation}
and
 \begin{equation}
	 \alpha^* = \frac{1}{N} \sum_{j=1}^{N} \alpha^*_j.
\end{equation}
Here $N$ is the number of exposures, and $\alpha$ and $\alpha_j$  are
the central wavelengths of the Br-$\gamma$ absorption line
in the full combined spectrum and in the $j$th partial spectra, respectively
\citep{12Wallpsa}.
Fig. \ref{fig:specBr-g16JK} shows four partial-data spectra (from 1st to 4th) for 2016.
We obtained $\sigma_{\mathrm{JK}} = 11.8$\,km/s for $s = 11$.

\begin{figure}[!ht]
  \begin{center}
    \includegraphics[width=0.6\textwidth,angle=0]{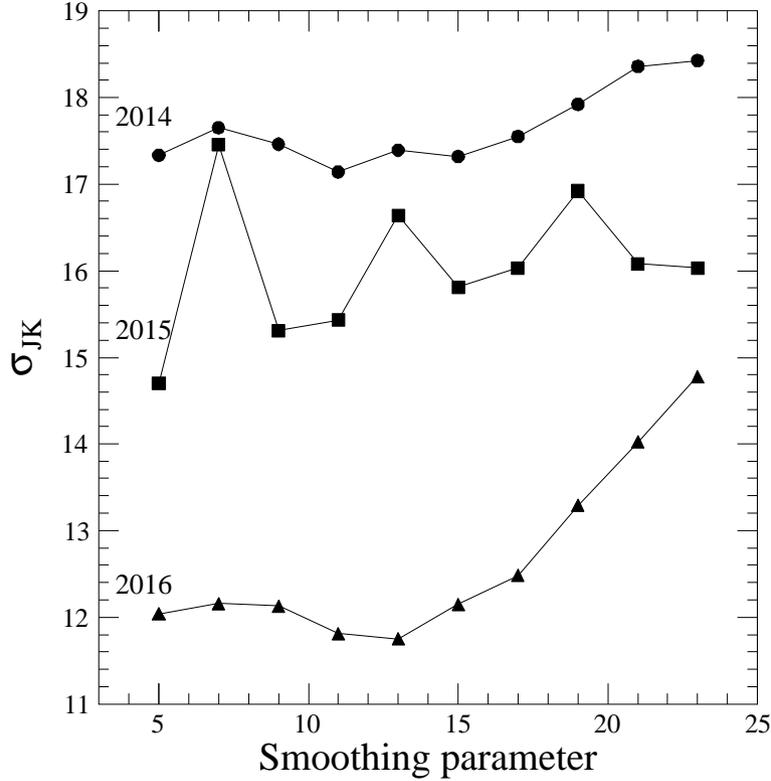}
  \end{center}
  \caption{
	Plot of $\sigma_{\mathrm{JK}}$ as a function of smoothing parameter $s$,
	for 2014 (circle), 2015 (squares), and 2016 (triangles).
  }
  \label{fig:SmoothSigJK}
\end{figure}

We have carried out the same analysis for different smoothing parameters.
The results are shown in Fig. \ref{fig:SmoothSigJK}.
For the 2016 data sets (triangles), when the smoothing parameter increases,
$\sigma_{\mathrm{JK}}$ slightly increases but decreases at $s = 11$,
and increases again for large $s$.
The minimum $\sigma_{\mathrm{JK}}$ appears at $s = 11 - 13$.
For 2015 data sets (squares in Fig. \ref{fig:SmoothSigJK}), 
the variation in $\sigma_{\mathrm{JK}}$ is larger than 2016 and 2014,
and there is no clear local minimum.
For 2014 (circles in Fig. \ref{fig:SmoothSigJK}),
we have found a local minimum at around $s = 9 - 15$.
As shown in Fig. \ref{fig:specBr-g16Smooth} and Table \ref{Tab:Brg_16_spar},
the standard deviations in RV$_{\mathrm{LSR}}$ for different smoothing parameter
with $s \leq 23$ are only a few km/s.
So in the following analysis, we will use results with $s = 11$.

The resultant RV$_{\mathrm{LSR}}$ and $\sigma_{\mathrm{JK}}$ for $s=11$
are listed in Table \ref{Tab:RV_Brg_s11}.
The shown jackknife RV$_{\mathrm{LSR}}$ are the mean of all the partial-data sets.
As shown in Table \ref{Tab:Brg_16_spar}, 
RV$_{\mathrm{LSR}}$ from the fitting of the all combined spectra with $s = 11$
are 497.3\,km/s, 876.5\,km/s, and 1108.9\,km/s
for 2014, 2015, and 2016, respectively.
The differences in RV$_{\mathrm{LSR}}$ from the jackknife analysis are very small.

\begin{table}[!ht]
\begin{center}
\caption{RV$_{\mathrm{LSR}}$ and $\sigma_{\mathrm{JK}}$ from the jackknife analysis for $s = 11$.
\label{Tab:RV_Brg_s11}} 
\begin{tabular}{ccccccc}
\hline
& \multicolumn{2}{c}{2014}  & \multicolumn{2}{c}{2015} & \multicolumn{2}{c}{2016}  \\ \hline
& RV$_{\mathrm{LSR}}$ & $\sigma_{\mathrm{JK}}$
& RV$_{\mathrm{LSR}}$ & $\sigma_{\mathrm{JK}}$ & RV$_{\mathrm{LSR}}$ & $\sigma_{\mathrm{JK}}$\\ \hline
jackknife & 497.0 & 17.1 & 876.7 & 15.4 & 1107.5 & 11.8 \\ 
combined & 497.3 & & 876.5 & & 1108.9 & \\
\hline
\end{tabular}
\end{center}
\end{table}

\subsection{Uncertainty in Wavelength Calibration}
\label{subsec:WaveCalib}

\subsubsection{Short Term Stability of Wavelength Calibration}

Our monitoring observations of S2 were carried out
for $\sim 4$\,hrs in 2014, $\sim 2$\,hrs in 2015, and 2 nights in 2016.
Any uncertainty from unstable wavelength calibrations
on scales of hours or days
are included in the uncertainties derived using the jackknife analysis.
However, it is important to constrain the absolute uncertainty
of the wavelength calibration.

In our S2 spectra, Br-$\gamma$ emission from the local, interstellar gas 
around S2 is seen.
The interstellar gas is ionized by UV radiation from high mass stars around Sgr\,A*.
Assuming that the RV of the local gas is constant,
the  Br-$\gamma$ emission lines can be used to examine
how accurate our wavelength calibrations are.
To this aim, we fitted the Br-$\gamma$ emission profile with a Gaussian function
for each exposure,
to understand the stability of the wavelength calibration.

\begin{figure}[!ht]
  \begin{center}
    \includegraphics[width=0.6\textwidth,angle=0]{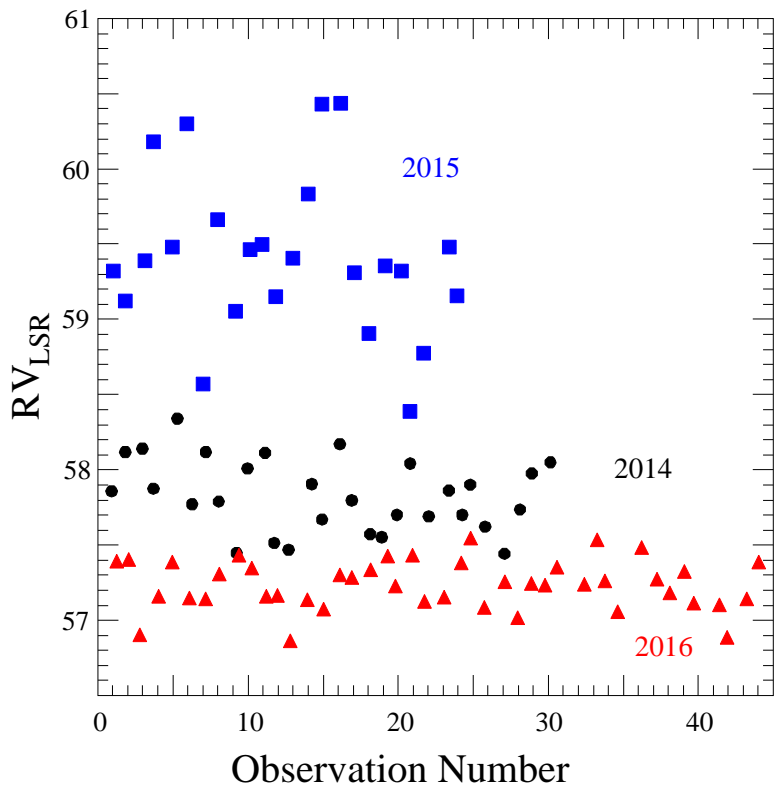}
  \end{center}
  \caption{
	RV$_{\mathrm{LSR}}$ of the Br-$\gamma$ emission line
	for each exposure as a function of the order of observations.
	RV$_{\mathrm{LSR}}$ for 2014, 2015, and 2016 are represented by
	black circles, blue squares, and red triangles, respectively.
  }
  \label{fig:Br-gEmiDist141516}
\end{figure}

We have found that
RV$_{\mathrm{LSR}}$ of the Br-$\gamma$ emission line is stable.
Fig. \ref{fig:Br-gEmiDist141516} shows
the RV$_{\mathrm{LSR}}$ for each exposure with $s = 1$
as a function of the order of the observations.
The standard deviations of RV$_{\mathrm{LSR}}$ 
in 2014 (black circles in Fig. \ref{fig:Br-gEmiDist141516}) is 0.24\,km/s.
That for 2015 is slightly larger, 0.53\,km/s, but still much smaller than $\sigma_{\mathrm{JK}}$.
The RV$_{\mathrm{LSR}}$ in 2016 are most stable, and the standard deviation is only 0.16\,km/s.
Therefore we conclude that our wavelength calibration is stable during one or two nights,
at least in the echelle order 26 where Br-$\gamma$ emission and absorption lines are found.
The short-term uncertainty in the wavelength calibration is negligible compared to 
the jackknife uncertainty.

\subsubsection{Long Term Stability of Wavelength Calibration}

The short term (from a few hours to 2 days) stability of the wavelength calibration
was examined in the previous section.
However, since our observations were carried out for three years,
an examination of a long-term (months or years) stability is necessary.

\begin{figure}[!ht]
  \begin{center}
    \includegraphics[width=0.6\textwidth,angle=0]{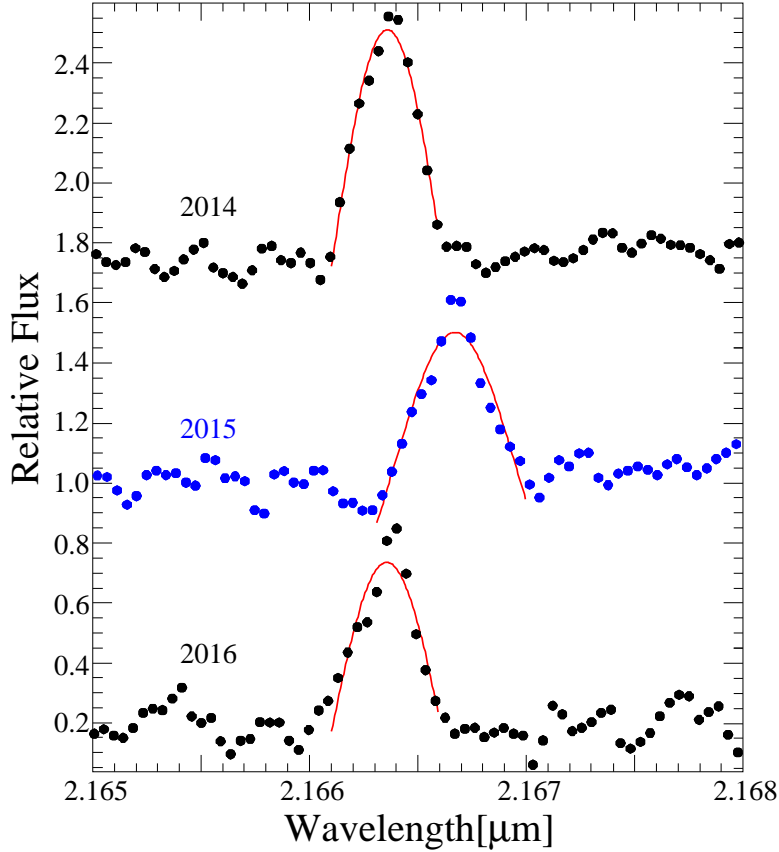}
  \end{center}
  \caption{
	Combined, full resolution ($s = 1$) spectra around the Br-$\gamma$ emission lines
	for 2014 (top), 2015 (middle), and 2016 (bottom).
	The emission profiles are fit with a Gaussian function.
	The difference of the wavelengths among the three spectra
	are mainly due to the motions of the Earth and the Solar system.
  }
  \label{fig:specBr-gEmi}
\end{figure}

We have derived RV$_{\mathrm{LSR}}$ of the Br-$\gamma$ emission line for each year,
by fitting the combined spectra ($s=1$) with a Gaussian function (Fig. \ref{fig:specBr-gEmi}).
The resultant peak wavelengths and RV$_{\mathrm{LSR}}$ are shown 
in Table \ref{Tab:RV_BrgEmission}.
The standard deviation of RV$_{\mathrm{LSR}}$ 
from 2014 to 2016 is $1.2\,$km/s.
If we simply use all the data points shown in Fig. \ref{fig:Br-gEmiDist141516},
the standard deviation is 0.93\,km/s.
These results suggest that the long-term, systematic uncertainty 
in our monitoring observations of S2 is small
compared to the statistical uncertainty derived by the jackknife analysis.

\begin{table}[!ht]
\begin{center}
\caption{Observed peak wavelength and RV$_{\mathrm{LSR}}$ of the Br-$\gamma$ emission lines.
\label{Tab:RV_BrgEmission}} 
\begin{tabular}{ccccccc}
\hline
year & $\lambda$ [$\mu$m] & RV$_{\mathrm{LSR}}$ [km/s] \\ \hline
2014 & 2.166360 & 57.8 \\
2015 & 2.166663 & 59.5 \\
2016 & 2.166357 & 57.2 \\ \hline
mean & & 58.2 \\
$\sigma$ & & 1.2 \\
\hline
\end{tabular}
\end{center}
\end{table}

\subsection{Partly Excluded Spectrum}

One of the difficulties in data reduction of NIR high-resolution spectroscopy
is the telluric correction.
Although we have tried to find better wavelength solutions,
we see uncorrected atmospheric lines, especially in the orders 28 and 29.
The number of the telluric lines is much smaller 
around the Br-$\gamma$ wavelength,
but weak, uncorrected lines could be included in the spectra we used to determine RV.

\begin{figure}[!h]
  \begin{center}
    \includegraphics[width=0.6\textwidth,angle=0]{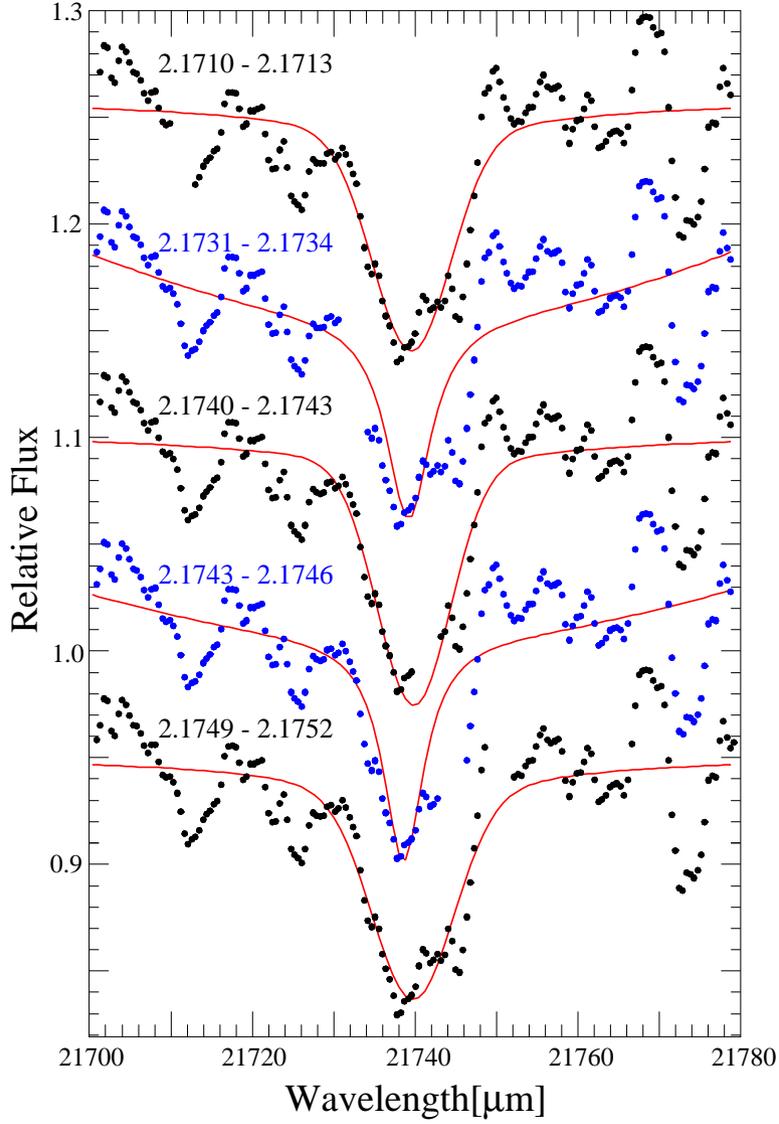}
  \end{center}
  \caption{
    2016 combined spectra with $s = 11$, but a part of the data points is excluded.
	From top to bottom, the wavelength ranges excluded are 
	$2.1710 - 2.1713\,\mu$m,
	$2.1731 - 2.1734\,\mu$m,
	$2.1740 - 2.1743\,\mu$m,
	$2.1743 - 2.1746\,\mu$m, and
	$2.1749 - 2.1752\,\mu$m.
	The Br-$\gamma$ features are fit with a Moffat function (red lines).
  }
  \label{fig:specBr-g16_partialSpec}
\end{figure}

We have examined how the central wavelength of the Br-$\gamma$ absorption line changes
when a part of the spectra is excluded.
In the case of 2016, we have made 20 spectra
without several data points in the fitting range of the Br-$\gamma$ line.
The width of the excluded wavelength range is $0.0003\,\mu$m  $= 3\,$\AA,
in which $6 -7$ data points are included.
Five of the partly excluded spectra are shown in Fig. \ref{fig:specBr-g16_partialSpec}.
The top one shows the combined 2016 spectra of $s = 11$,
without data points  in the range of 
$2.1710\,\mu$m $< \lambda < 2.1713\,\mu$m.

Table \ref{Tab:Brg_16_partialSpec} shows the results of the fitting of the Br-$\gamma$ line
for the partly excluded spectra in 2016.
When a part of the wing region is excluded, the shift of the central wavelength is very small.
On the other hand, when a part of the core region is excluded, 
the peak wavelength shifts slightly larger.

We can see substructures in the core region of the Br-$\gamma$ absorption line, even with $s = 11$.
There is likely to be some peaks in the profile, although whether the substructures are real or not
is still not clear due to the low S/N ratio of our spectra.
When a substructure is in the excluded region, 
the shape of the fitting curve changes, and the resultant central wavelength shifts.
However, as shown in Table \ref{Tab:Brg_16_partialSpec},
the standard deviation of the RV$_{\mathrm{LSR}}$ is small, 3.6\,km/s.
Even if we focus on the spectra with the excluded region near the spectral core,
$2.173\,\mu$m $\lesssim \lambda \lesssim 2.175\,\mu$m,
the standard deviation appears to be $\sim 6$\,km/s.

\begin{table}[!ht]
\begin{center}
\caption{Wavelength of Br-$\gamma$ and RV$_{\mathrm{LSR}}$ for partially excluded spectra for $s = 11$. 
\label{Tab:Brg_16_partialSpec}} 
\begin{tabular}{ccccccc}
\hline
\hline
excluded range [$\mu$\,m]  & $\lambda$ [$\mu$\,m] & RV$_{\mathrm{LSR}}$ [km/s] \\ \hline
$2.1710 - 2.1713$ & 2.173955  & 1108.8 \\
$2.1713 - 2.1716$ & 2.173955  & 1108.8 \\
$2.1716 - 2.1719$ & 2.173956  & 1108.9 \\
$2.1719 - 2.1722$ & 2.173956  & 1108.9 \\
$2.1722 - 2.1725$ & 2.173957  & 1109.1 \\
$2.1725 - 2.1728$ & 2.173964  & 1110.1 \\
$2.1728 - 2.1731$ & 2.173957  & 1109.2 \\
$2.1731 - 2.1734$ & 2.173918  & 1103.7 \\
$2.1734 - 2.1737$ & 2.173965  & 1110.3 \\
$2.1737 - 2.1740$ & 2.173960  & 1109.6 \\
$2.1740 - 2.1743$ & 2.173983  & 1112.7 \\
$2.1743 - 2.1746$ & 2.173856  & 1095.1 \\
$2.1746 - 2.1749$ & 2.173956  & 1109.1 \\
$2.1749 - 2.1752$ & 2.173974  & 1111.5 \\
$2.1752 - 2.1755$ & 2.173957  & 1109.1 \\
$2.1755 - 2.1758$ & 2.173957  & 1109.2 \\
$2.1758 - 2.1761$ & 2.173956  & 1108.9 \\
$2.1761 - 2.1764$ & 2.173955  & 1108.9 \\
$2.1764 - 2.1767$ & 2.173955  & 1108.9 \\
$2.1767 - 2.1770$ & 2.173960  & 1109.5 \\\hline
mean &  2.173952 & 1108.5 \\
$\sigma$ & 0.000026 & 3.6 \\
\hline
\end{tabular}
\end{center}
\end{table}

\begin{figure}[!h]
  \begin{center}
    \includegraphics[width=0.35\textwidth,angle=-90]{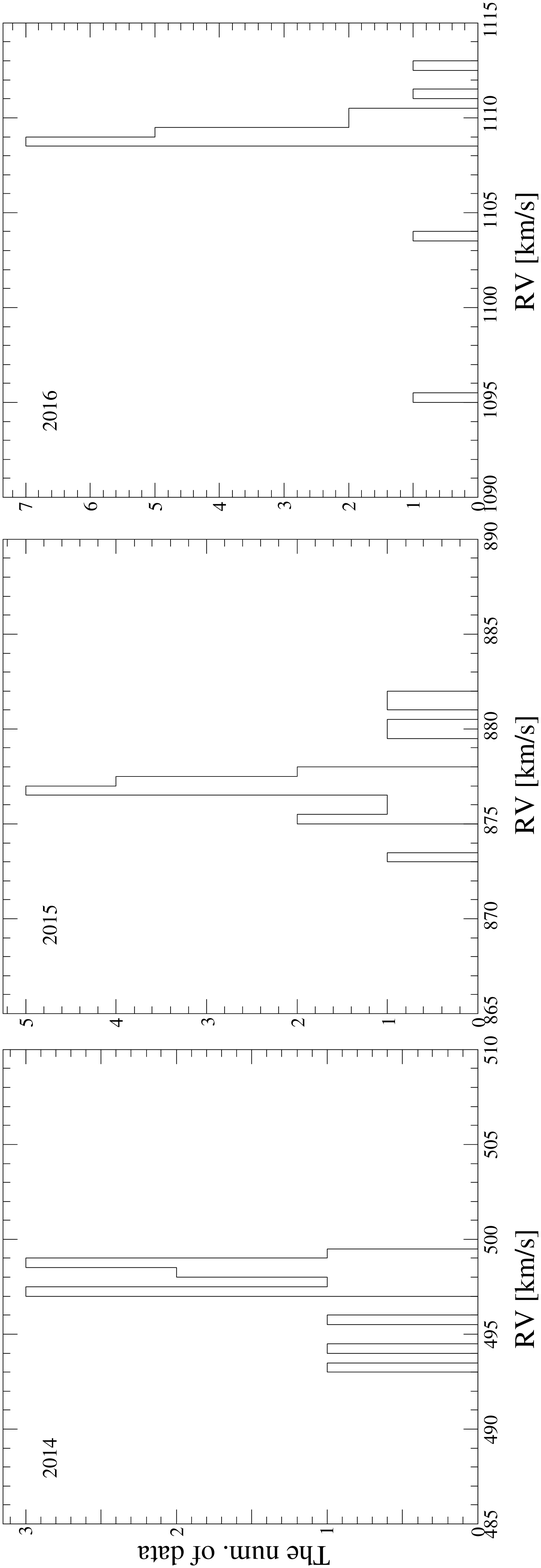}
  \end{center}
  \vspace{0.1cm}
  \caption{
	Distributions of RV$_{\mathrm{LSR}}$ for partly excluded spectra
	for 2014 (left), 2015 (middle), and 2016 (right).
	The standard deviations of RV$_{\mathrm{LSR}}$ are
	1.9, 2.0, and 3.6\,km/s for 2014, 2015, and 2016, respectively.
  }
  \label{fig:RV141516_partialSpec}
\end{figure}

A similar analysis was done for 2014 and 2015.
The fitting wavelength range for the 2014 spectrum is smaller than 2016.
We have made 13 spectra excluding $6 - 7$ consecutive data points ($= 3$\,\AA)
in the range of $2.1680\,\mu$m $< \lambda < 2.1720\,\mu$m for the 2014 data sets.
The result of the fitting of the 13 spectra is shown in Fig. \ref{fig:RV141516_partialSpec},
left panel.
The standard deviation of RV$_{\mathrm{LSR}}$ is 1.9\,km/s.

We have made 20 spectra for the 2015 data sets, and carried out the same analysis.
The result is shown in the middle panel in Fig. \ref{fig:RV141516_partialSpec},
and the standard deviation of RV$_{\mathrm{LSR}}$ is 2.0\,km/s.

As shown above, even if a part of the spectra is excluded,
the resultant RV$_{\mathrm{LSR}}$ changes only a few km/s.
Hence we conclude that the RV$_{\mathrm{LSR}}$ of S2
derived from the Br-$\gamma$ absorption is not 
strongly affected by the incomplete telluric correction.

\section{Discussion}

\subsection{Summary of RV$_{\mathrm{LSR}}$ and Uncertainties}

The obtained RV$_{\mathrm{LSR}}$ and uncertainties discussed above 
are summarized in Table \ref{Tab:ErrorBudget}.
We use the jackknife $\sigma_{\mathrm{JK}}$ as statistical uncertainties.
As total systematic uncertainties, we quadratically added the uncertainties
from the spectrum smoothing, 
the long-term stability of the wavelength calibration,
and the partly excluded spectrum analysis. 
The uncertainty in the short-term stability of the wavelength calibration
is included in the uncertainties from the jackknife analysis,
and thus we do not add them separately.

\begin{table}[!ht]
\begin{center}
\caption{RV$_{\mathrm{LSR}}$ Error Budget in the unit of km/s.
\label{Tab:ErrorBudget}} 
\begin{tabular}{cccccccccc}
\hline
& RV$_{\mathrm{LSR}}$ & Statistic ($\sigma_{\mathrm{JK}}$) & \multicolumn{3}{c}{Systematic} & total$_{\mathrm{sys}}$ & total uncertainty \\ 
& & & smoothing & $\lambda$(long) & partial & \\ \hline
2014 &  497 & 17.1 & 1.5 &1.2 & 1.9 & 2.7 & 17.3 \\
2015 &  877 & 15.4 & 2.7 &1.2 & 2.0 & 3.6 & 15.8 \\
2016 & 1108 & 11.8 & 1.9 &1.2 & 3.6 & 4.2 & 12.5 \\
\hline
\end{tabular}
\end{center}
\end{table}

\subsection{He\,I absorption lines at 2.1126\,$\mu$m}

\begin{figure}[!ht]
  \begin{center}
    \includegraphics[width=0.8\textwidth,angle=0]{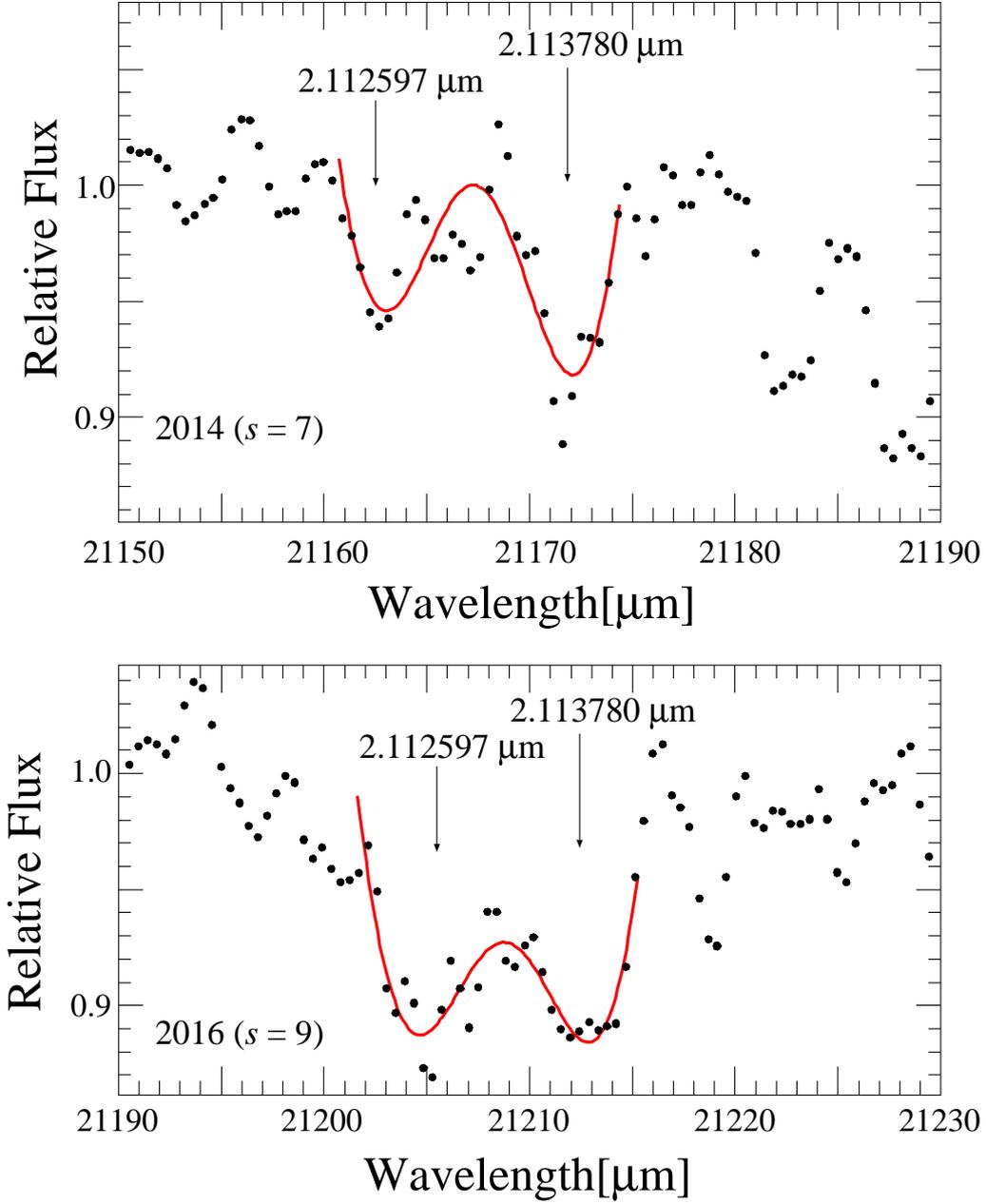}
  \end{center}
  \caption{
  	2014 (top) and 2016 (bottom) combined spectra around He\,I lines.
	The smoothing parameters for the shown spectra are
	$s = 7$ and $s = 9$ for 2014 and 2016, respectively.
	The positions of the 2.112597\,$\mu$m and 2.113780\,$\mu$m lines
	are indicated by arrows.
	The profiles are fit with a double Gaussian function to determine the peak wavelengths (red curves).
  }\label{fig:spec1416HeI}
\end{figure}

Although the S/N ratio is low, 
He\,I absorption lines at 2.112597\,$\mu$m and 2.113780\,$\mu$m
are detected in our 2014 and 2016 spectra (Fig. \ref{fig:spec1416HeI}).
We use the He\,I lines to check the reliability of RV$_{\mathrm{LSR}}$
and their uncertainties from the Br-$\gamma$ line.

We fitted the two He\,I lines with a double Gaussian function
(red curves in Fig. \ref{fig:spec1416HeI}).
The two He\,I lines are so close that we cannot resolve them
with the smoothing parameters of $s \geq 11$.
The wavelength difference between the Gaussian peaks are fixed to be
$2.1137800 - 2.1125965 =  0.00011835\,\mu$m,
and the peak wavelength of the He\,I 2.112597\,$\mu$m line is set to be a free parameter
in the fitting procedure.
The smoothing parameters for the spectra in Fig. \ref{fig:spec1416HeI}
are $s = 7$ and $9$ for 2014 and 2016, respectively.
We cannot find a clear He\,I feature for the 2015 combined spectrum.

We have derived RV$_{\mathrm{LSR}}$ from the measured
He\,I 2.112597\,$\mu$m line peak wavelengths.
The peak wavelengths and corresponding RV$_{\mathrm{LSR}}$ are 
shown in Table \ref{Tab:RV_HeI}.
The obtained He\,I RV$_{\mathrm{LSR}}$ for 2014, 529.6\,km/s, is larger than 
that for the Br-$\gamma$ line ($\approx 497\,$km/s),
although the difference is less than $2\,\sigma$.
The He\,I RV$_{\mathrm{LSR}}$ for 2016 is in good agreement
with that from the Br-$\gamma$ line ($\approx 1108\,$km/s),
and the difference is smaller than $1\,\sigma$ uncertainty of 12.5\,km/s.
We thus conclude that the RV$_{\mathrm{LSR}}$ values 
from the Br-$\gamma$ and He\,I lines are consistent with each other.
If we simply calculate means and standard deviations of the mean from the Br-$\gamma$ and He\,I lines,
the derived RV$_{\mathrm{LSR}}$ are  $513.3 \pm 16.2$\,km/s and $1113.6 \pm 5.4$\,km/s
for 2014 and 2016, respectively.

\begin{table}[!ht]
\begin{center}
\caption{Observed He\,I 2.112597\,$\mu$m wavelengths and RV$_{\mathrm{LSR}}$.
\label{Tab:RV_HeI}} 
\begin{tabular}{ccccccc}
\hline
& 2014 ($s = 7$)  & 2016 ($s = 9$)  \\ \hline
observed $\lambda$ [$\mu$m] & 2.116156 & 2.120309\\ 
RV$_{\mathrm{LSR}}$ [km/s] & 529.6 & 1118.9\\ \hline
mean RV$_{\mathrm{LSR}}$ [km/s] $^{\mathrm{a}}$ & $513.3 \pm 16.2 $ & $113.6 \pm 5.4$  \\ \hline
\end{tabular}
\end{center}
(a) Mean and standard error of the mean of RV$_{\mathrm{LSR}}$ from Br-$\gamma$ and He\,I lines.
\end{table}

\subsection{Wider Br-$\gamma$ Absorption Profile in 2014}

As shown in Fig. \ref{fig:specBr-g141516Allfit}, the Br-$\gamma$ profile in the 2014 spectrum
is wider than those for 2015 and 2016.
One possible reason for the wider profile is an imperfect correction of the telluric absorption.
There is a strong atmospheric absorption line at $\lambda \approx 2.1687\,\mu$m,
and this absorption profile is within the 2014 Br-$\gamma$ profile.
The telluric profile is well corrected in 2016, but there seems to be residual in 2015.
Hence the 2014 Br-$\gamma$ profile could be extended to the bluer wavelength
due to the residual of the telluric line at $\approx 2.1687\,\mu$m,
and this could lead to the difference in RV$_{\mathrm{LSR}}$ 
between the Br-$\gamma$ and He\,I lines.

To check instrumental effects for the width of the Br-$\gamma$ profile,
we have compared the widths of the Br-$\gamma$ emission line at $\sim 2.1665\,\mu$m
(see \S \ref{subsec:WaveCalib}).
As shown in Fig. \ref{fig:specBr-gEmi},
the Br-$\gamma$ emissions were fit with a Gaussian function.
The obtained Gaussian sigmas are 
2.4\,\AA, 3.3\,\AA, and $2.8\,\mathrm{\AA}$  
in 2014, 2015, and 2016, respectively.
The difference in sigma could be explained by the difference of the observation modes;
In 2014 and 2016, we could use the LGS system, which makes an AO guide star
at a closer position to S2 than natural guide stars.
In 2015, we could not use the LGS system, and thus the spatial resolution was worse than other 2 epochs.
It could lead to observations of ionized gas at larger region.
However, the Gaussian sigma for 2014 is smaller than 2015 and 2016,
and it cannot explain the wider Br-$\gamma$ profile in the 2014 spectrum.

The wider profile might be explained by intrinsic properties of S2.
Possible origin is 
a change of the direction of the S2's rotation axis, or binarity of S2.
However, no flux variation due to a binary eclipse
has been detected \citep{Rafelski07ApJ, Ghez08PM, Gillessen09PM}.
We continue to investigate the intrinsic properties of S2 in upcoming monitoring observations.

\subsection{S2 RV Curve since 2000}

\begin{figure}[!h]
  \begin{center}
    \includegraphics[width=0.8\textwidth,angle=0]{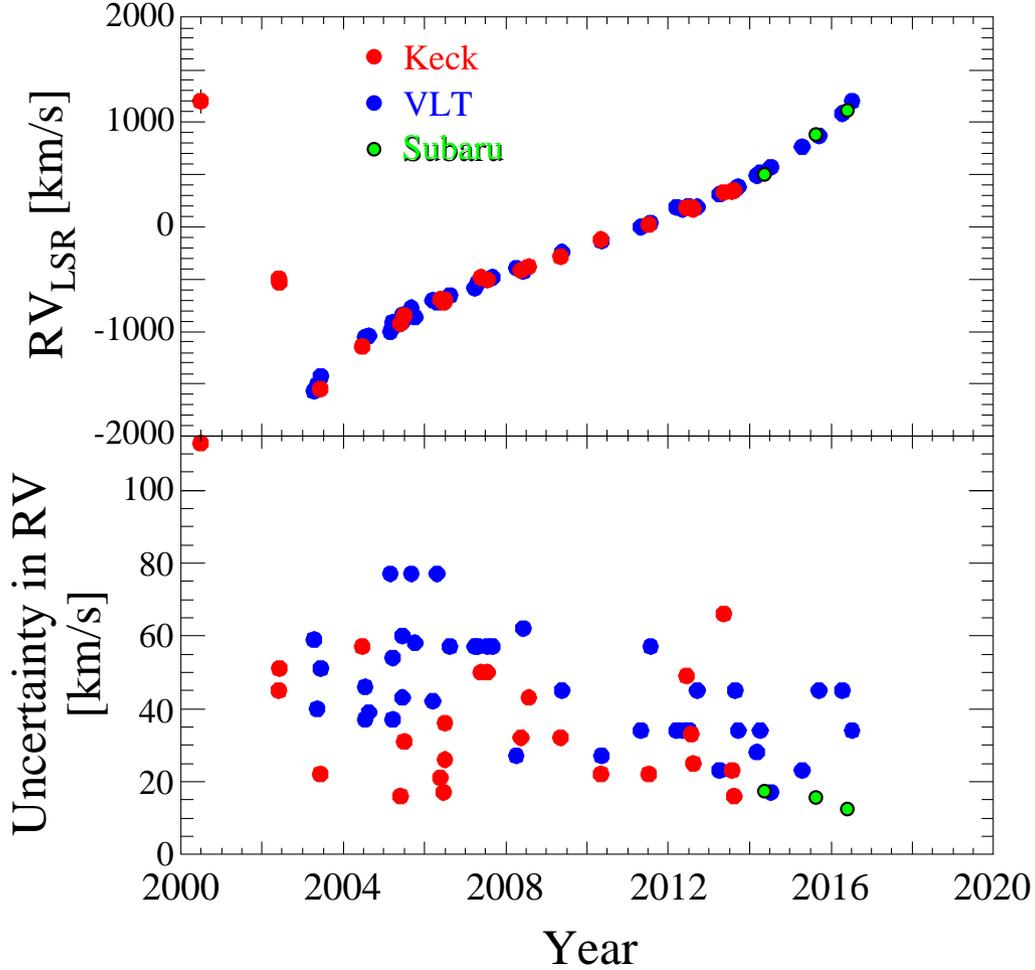}
  \end{center}
  \caption{
    Measured RV$_{\mathrm{LSR}}$ (top) and uncertainties in RV$_{\mathrm{LSR}}$ (bottom)
	as a function of time.
	Our results using the Subaru telescope are shown by green circles.
	The red data, using the Keck telescopes, are from \citet{Boehle16ApJ},
	and the blue data, using VLT, are from \citet{17GillessenApJ}.
  }
  \label{fig:RVplotS2}
\end{figure}

In Fig. \ref{fig:RVplotS2}, we show the plots of RV$_{\mathrm{LSR}}$ 
and uncertainties in RV$_{\mathrm{LSR}}$ as a function of time,
combined with the past observations using the Keck telescope \citep{Boehle16ApJ}
and VLT \citep{17GillessenApJ}.
In the past measurements, the uncertainties mainly range from $\sim 20$\,km/s to $\sim 60$\,km/s. 
The mean RV$_{\mathrm{LSR}}$ uncertainty is 34\,km/s since 2010,
and the best measurement was in 2013 with an uncertainty of 16\,km/s
using Keck/OSIRIS with a total exposure time of $\approx 7.8$\,hrs.

As shown in Fig. \ref{fig:RVplotS2}, our RV$_{\mathrm{LSR}}$ uncertainties are stable,
and the mean of them are smaller than that of the past measurements.
One of the reasons why the uncertainties of our measurements are smaller than the past ones 
is probably an accuracy of the wavelength calibration.
We have used the atmospheric OH emission lines for the calibration.
They have narrow features even in our full resolution spectra,
and it means that the peak wavelengths of the OH lines can be determined with a better accuracy
if we use higher-spectral resolution spectrograph.
As shown in \S \ref{sec:RVandUncertainty},
the uncertainties of the wavelength calibration on the scales of hours or days
are less than $\sim 0.5$\,km/s in 2014, 2015, and 2016.
The long-term uncertainty in the calibration is also small
with a standard deviation of 1.2\,km/s.
In the past observations, 
the uncertainties in the wavelength calibration were $\sim 9$\,km/s
\citep{Ghez03ApJ, Ghez08PM} for S2/S0-2,
although it was estimated to be in the order of $2-3$\,km/s using VLT \citep{Gillessen09PM}.

Another reason is that our spectral resolution is high enough to separate
the Br-$\gamma$ line profile from nearby He\,I absorption lines.
As shown in \citet{08MartinsApJ} and \citet{Habibi17arXiv},
in the medium-resolution spectroscopy,
the He\,I lines at $\sim 2.162\,\mu$m are in the wing of the Br-$\gamma$ profile.
However, in our spectra for 2015 and 2016, 
the He\,I lines are clearly separated from the Br-$\gamma$ line,
and are at the edge or out of our fitting range.
Hence the He\,I lines do not affect the peak wavelength measurements 
of the Br-$\gamma$ line.

\subsection{RV Measurements of S2 in 2018}

\begin{figure}[!h]
  \begin{center}
    \includegraphics[width=0.6\textwidth,angle=0]{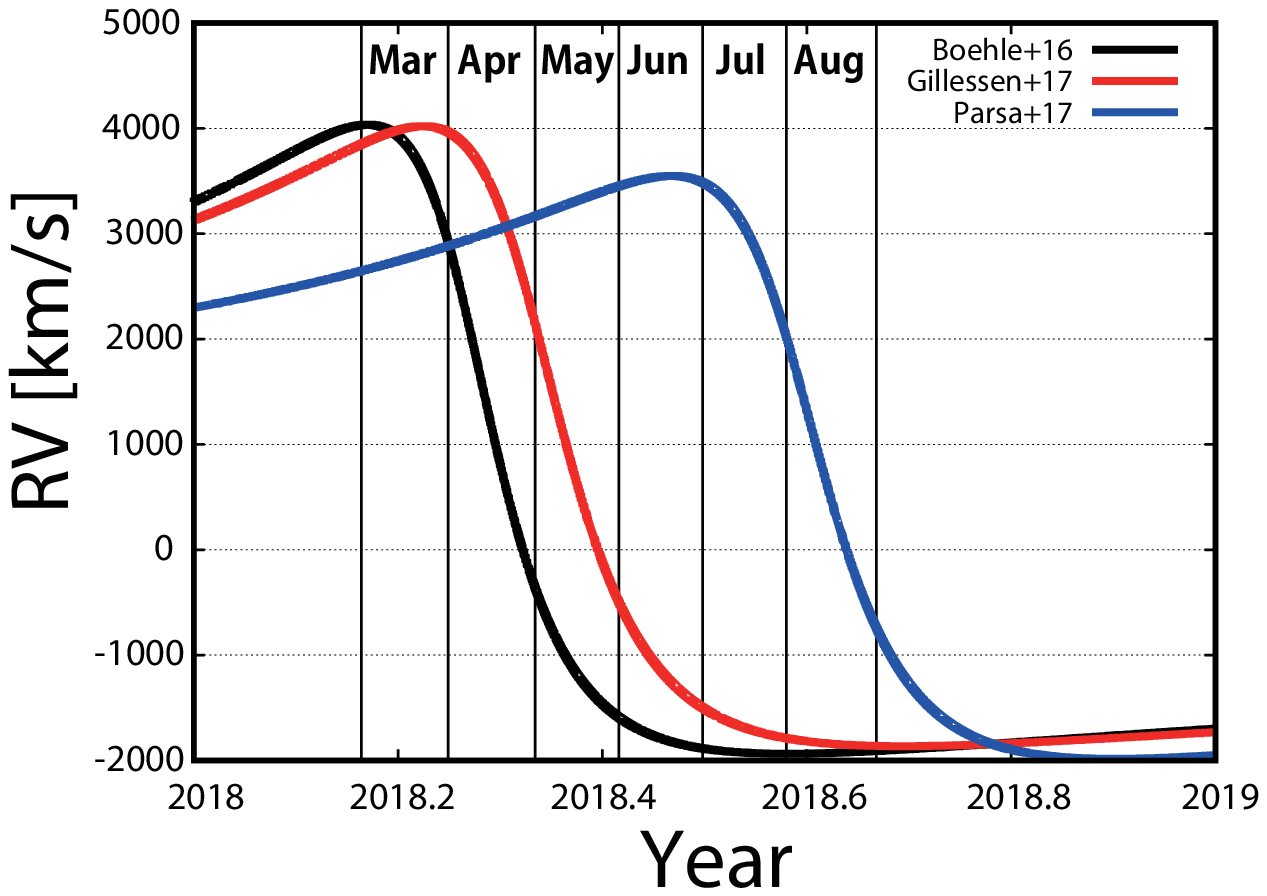}
  \end{center}
  \caption{
	Expected S2 RV curves in 2018, using the parameters
	in \citet{Boehle16ApJ} (black curve),
	\citet{17GillessenApJ} (red curve), and \citet{Parsa17Aph} (blue curve).
  }
  \label{fig:RVplotS2_2018}
\end{figure}

To show the importance of RV measurements of S2 in 2018,
in Fig. \ref{fig:RVplotS2_2018},
we compare the expected RV curves using orbital parameters derived by
the most recent works:
\citet{Boehle16ApJ}; \citet{17GillessenApJ}; and \citet{Parsa17Aph}.
Here the curves in Fig.~\ref{fig:RVplotS2_2018} show 
the expected RV curves from pure Keplerian motions,
where no relativistic effect is included. 

A number of astrometric observations with the Keck telescope and NTT/VLT,
and careful data analysis have provided us with
strong constraints on the mass $M_{\mathrm{Sgr\,A*}}$ and the distance to the Galactic SMBH. 
The amount of mass concentrated around Sgr\,A* has been estimated
with an uncertainty of $3 - 4$\,\% \citep{Boehle16ApJ,17GillessenApJ}.
However, the predicted next pericenter passages are
$2018.29 \pm 0.04$ \citep{Boehle16ApJ},
$2018.35 \pm 0.02$ \citep{17GillessenApJ}, and
$2018.59 \pm 0.21$ \citep{Parsa17Aph};
the difference is as large as 0.3\,yr $\approx 110$\,days.
Although this is only a few \% uncertainty of the S2's orbital period,
this is still large for a detailed, appropriate planning of observations in 2018.
As one can see, the differences among the expected RV curves
could be more than 1,000\,km/s in 2018.
Frequent spectroscopic measurements of S2 in 2018,
especially during the steep decline phase of RV,
will allow us to reduce the uncertainty in the orbital period of S2.

\begin{figure}[!h]
  \begin{center}
    \includegraphics[width=0.9\textwidth,angle=0]{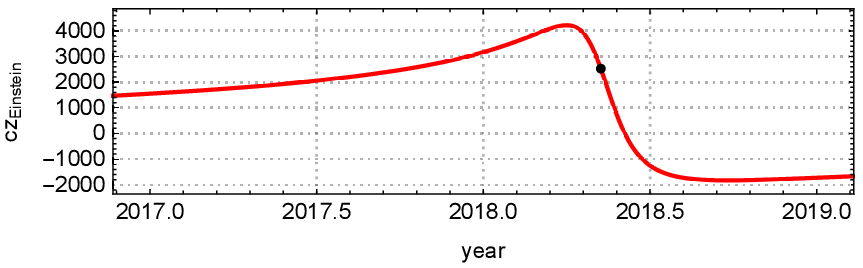}
    \includegraphics[width=0.9\textwidth,angle=0]{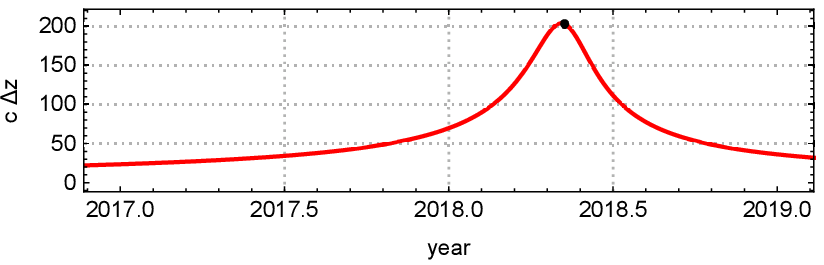}
  \end{center}
  \caption{
    Top: Time evolution of $cz_{\mathrm{Einstein}}$, 
    where the BH mass and S2's orbital parameters are 
    the best-fit values given by \citet{17GillessenApJ}.
    We assumed that the spin direction is pointing the Galactic south 
    and the spin magnitude is 0.98 $M_{\mathrm{Sgr\,A*}}$
    (98\% of the theoretically allowed maximum value).
    Bottom: Time evolution of the general relativistic effect $c \Delta z$.
    The timing of the pericenter passage is shown by black dot.
  }
  \label{fig:czPlotS2}
\end{figure}

Next, let us discuss the expectation for the detection of 
the PN effects in RV measurements. 
As noted in \S \ref{sec:intro}, 
the relation between the redshift of photons coming from S2, $z$, 
and the radial velocity of S2, $v_{\mathrm{S2}}$, 
is complicated and our observable quantity is not exactly
$v_{\mathrm{S2}}$ but $z$ (see equation (\ref{eq:eq1})).
Thus, we define the GR effect measured in spectroscopic observations as 
\begin{equation}
c \Delta z = cz_{\mathrm{Einstein}} - cz_{\mathrm{Newton}},
\end{equation}
where $cz_{\mathrm{Einstein}}$ is the redshift estimated by GR 
(S2 motion and photon propagation in the rotating BH spacetime),
and $cz_{\mathrm{Newton}}$ is the redshift by the Newtonian mechanics 
(S2 motion in the point mass Newtonian gravitational potential). 
Our estimation of $cz_{\mathrm{Einstein}}$ and $c \Delta z$ are shown in Fig. \ref{fig:czPlotS2}. 
In our theoretical calculation, the mass of Sgr\,A*, $M_{\mathrm{Sgr\,A*}}$,
and the orbital elements of S2,
which provide us with the initial condition for the S2 motion, are taken from \citet{17GillessenApJ}.
Fig. \ref{fig:czPlotS2} shows that 
$c \Delta z$ will become a few 10 km/s in the latter half of 2017,
 reach about 200 km/s near the next pericenter passage in 2018, 
 and fall to a few 10 km/s at the end of 2018. 
Our observational uncertainties in $cz $ of $12 - 17\,$km/s, 
enable us to detect the GR effects
in the spectroscopic measurements within the next one and a half years.

Note that the so-called PN expansion of $cz_{\mathrm{Einstein}}$ 
can be expressed as a series expansion (polynomial in $1/r$), 
\begin{equation}
\label{eq:cz}
	cz_{\mathrm{Einstein}} = cz_{\mathrm{Newton}} + \mathrm{1st~PN~term} + \mathrm{higher~order~PN~terms} ,
\end{equation}
where the order of the small parameter of this expansion is estimated by 
$G M_{\mathrm{Sgr\,A*}}/(c^2 r_{\mathrm{peri}}) \sim 10^{-3}$, 
where $r_{\mathrm{peri}} \simeq 121$ AU is the pericenter distance of S2,
and $G$ is the gravitational constant. 
The 1st PN term for the S2 dynamics consists mainly of 
the kinematic Doppler effect and the gravitational redshift.
Each of these two effects is estimated to be about 100 km/s ($\sim c \times 10^{-3}$) 
and the total of the 1st PN terms becomes 200 km/s near the pericenter passage 
(Fig. \ref{fig:czPlotS2}, bottom panel). 
This 1st PN effect includes the effect of the BH mass but not the effect of the BH spin. 
The 2nd order PN effect in Equ. (\ref{eq:cz}), 
which includes the effect of the BH spin, is typically about $200 \times 10^{-3} \sim 0.2$ km/s. 
Therefore the GR effects we can detect with Subaru/IRCS is the 1st PN order effects.

\begin{figure}[!h]
  \begin{center}
    \includegraphics[width=0.8\textwidth,angle=0]{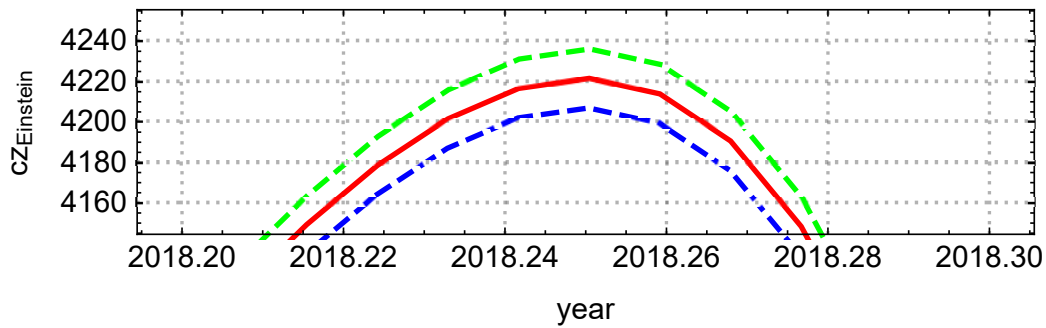}
  \end{center}
  \caption{
	Enlarged graph of $cz_{\mathrm{Einstein}}$ near the maximum value at 2018.2498. 
	Red curve is $cz_{\mathrm{Einstein}}$ estimated with the best-fit value of 
	$M_{\mathrm{Sgr\,A*}}$ given by \citet{17GillessenApJ}.
	Green and blue dashed curves are estimated 
    with 1\% larger and smaller SMBH masses, respectively,
	than the best-fit value. 
  }
  \label{fig:MassVariance}
\end{figure}

Fig. \ref{fig:MassVariance} shows the time evolution of $cz_{\mathrm{Einstein}}$ near its peak, 
assuming three different BH masses:
the best-fit value by \citet{17GillessenApJ} (red curve);
and 1\% larger (green dashed curve) and smaller (blue) than the best value.
The BH mass difference of 1\% makes $\sim 17$\,km/s shifts in the RV peaks.
The difference is almost the same with the parameters derived by \citet{Boehle16ApJ}.
These are almost the same amplitudes as the RV uncertainties we obtained with Subaru/IRCS.
Note that the measurements of $cz$ is almost independent of the distance 
from us to the Galactic center, $R_{\mathrm{GC}}$, 
because spectroscopic measurements do not strongly depend on 
the measurements of visible angle between the S2 and Sgr\,A*, 
and we do not need the value of $R_{\mathrm{GC}}$ to determine the value of $cz$. 
In the astrometric measurements, the degeneracy between $R_{\mathrm{GC}}$
and the mass of Sgr\,A* is a source of uncertainty
(e.g., Equ. (9) in \cite{17GillessenApJ}).
Hence, a combination of astrometric measurements and accurate spectroscopic measurements,
which is almost insensitive to $R_{\mathrm{GC}}$,
is expected to decrease the uncertainty of the mass of Sgr\,A*.

\section{Conclusion}

We have carried out near-infrared, high resolution spectroscopic observations
of S2 using Subaru/IRCS from 2014 to 2016.
The radial velocities of S2 were determined using the Br-$\gamma$ absorption line.
The total uncertainties in the radial velocity measurements are
17.3\,km/s, 15.8\,km/s, and 12.5\,km/s for 2014, 2015, and 2016, respectively.
We have confirmed the long-term stability of our radial velocity monitoring observations.
The uncertainties are smaller than those in the past, medium resolution spectroscopies,
and small enough to detect post-Newtonian effects in 2018.

\section{Acknowledgement} 
We thank the Subaru Telescope staff for the support for our observations.
This work was supported by JSPS KAKENHI, 
Grant-in-Aid for Young Scientists (A) 25707012,
Grant-in-Aid for Challenging Exploratory Research 15K13463,
H. S. was supproted by KAKENHI Grant-in-Aid for Challenging Exploratory Research 26610050.
Y. T. was supproted by KAKENHI Grant-in-Aid for Young Scientists (B) 26800150.
M.T. was supported by KAKENHI Grant Number 17K05439 
and DAIKO FOUNDATION.



\end{document}